\documentclass[a4paper,11pt]{article}
\usepackage{jheppub}

\pdfoutput=1
\usepackage{amsmath,amssymb,tabularx,graphicx,cancel,subcaption,xspace,flafter,ifpdf,todonotes,slashed,hyperref,cleveref,booktabs}
\usepackage{placeins}
\usepackage[force]{feynmp-auto}
\graphicspath{{figures/}}
\def\blfootnote{\xdef\@thefnmark{}\@footnotetext}

\DeclareMathOperator{\bucket}{bucket}
\DeclareMathOperator{\sgn}{sgn}

\begin{document}

\title{Unbiased Elimination of Negative Weights in Monte Carlo Samples
}

\author[a]{Jeppe~R.~Andersen}
\affiliation[a]{Institute for Particle Physics Phenomenology, Department of Physics, University of Durham, Durham, DH1 3LE, UK}

\author[b]{and Andreas Maier}
\affiliation[b]{Deutsches Elektronen-Synchrotron DESY, Platanenallee 6,
  15738 Zeuthen, Germany}

\abstract{
  We propose a novel method for the elimination of negative Monte
Carlo event weights. The method is process-agnostic, independent of
any analysis, and preserves all physical observables. We demonstrate
the overall performance and systematic improvement with increasing
event sample size, based on predictions for the production of a W boson with
two jets calculated at next-to-leading order perturbation theory.
}

\preprint{\begin{minipage}[t]{\widthof{MCNET-21-14,}}
    DCPT/21/54\\
    DESY 21-135\\
    IPPP/21/27\\
    MCNET-21-14\\
    SAGEX-21-29
  \end{minipage}}
\vspace*{10ex}

\maketitle

\section{Introduction}
\label{sec:intro}

The LHC physics programme has entered an era of precision
measurements. Such measurements demand equally precise theory
predictions. To match this requirement, it is most often necessary to
include at least next-to-leading order (NLO) perturbative
corrections. For a number of processes, even higher fixed-order
corrections, i.e. NNLO and NNNLO, have to be taken into account. In
order to consistently combine virtual corrections, usually computed in
$d=4-2\epsilon$ dimensions, and real corrections, obtained via
numerical integration over four-dimensional momenta, one introduces
subtraction terms which render each component finite, see for
example~\cite{Catani:1996vz,Catani:2002hc}.

Since one is usually interested in distributions, additional effects
from parton showers and hadronisation have to be accounted
for. Methods such as MC@NLO~\cite{Frixione:2002ik} and
POWHEG~\cite{Nason:2006hfa} can be employed for matching parton
showers and NLO fixed-order predictions. Depending on the phase-space
region of interest, further resummation may be required. Examples
include transverse momentum~\cite{Collins:1984kg} and high-energy resummation~\cite{Fadin:1975cb,Balitsky:1978ic}.
High-accuracy predictions for multi-jet observables are then obtained by
merging exclusive event samples using e.g. the
MEPS@NLO~\cite{Hoeche:2012yf} or
UNLOPS~\cite{Lonnblad:2012ix,Lonnblad:2012ng} approaches. Finally, the
detector response to the generated events is
simulated~\cite{GEANT4:2002zbu}.

Thanks to continuous improvements, modern general purpose event
generators~\cite{Alwall:2014hca,Bellm:2015jjp,Sjostrand:2014zea,Bothmann:2019yzt}
render the inclusion of many of these corrections feasible for a large
range of scattering processes. However, the improved accuracy comes at
a significant cost in computing time. What is more, each of the
individual calculational steps outlined above also increases the cost
of all subsequent steps. The reason for this is that typically a
number of negative-weight counterevents are generated, for instance
for unitarisation or in order to prevent double counting. This implies
that a much larger number of events has to be processed to reach the
same statistical significance as for the case of purely
positive-weight event samples. This problem is particularly pronounced
in the final detector simulation step, which can take hours of CPU
time for each event.

In general, the number of events that have to be forwarded to the
detector simulation can be reduced by unweighting the event
sample. Taking $W$ as the largest absolute event weight in the sample,
an event with weight $w_i$ is rejected with probability $\mathbb{P} =
1 - |w_i|/W$ and otherwise assigned the new weight $w_i \to \sgn(w_i)
W$. If all event weights are positive, an unweighted sample of $N$
events has uniform weight $w_i=W$, which is ideal in the sense that the
Monte Carlo uncertainty estimate for the cross section, $\langle
\sigma \rangle = \sqrt{\sum_{i=1}^N w_i^2}$, is minimal among all
samples with $N$ events and the same predicted cross section $\sigma =
\sum_{i=1}^N w_i$. Conversely, a given precision goal can be reached
with the smallest number of events.

This ideal case of completely uniform weights is no longer achieved as
soon as negative weights appear. For a fractional negative weight contribution
\begin{equation}
  \label{eq:r_-}
  r_- = -\frac{\sum_{w_i < 0} w_i}{\sum_i |w_i|},
\end{equation}
the number of events required for a given precision goal increases to
\begin{equation}
  \label{eq:N_r_-}
  N(r_-) = \frac{N(0)}{(1 - 2r_-)^2}
\end{equation}
relative to the required number $N(0)$ in the absence of negative
weights, see also~\cite{Danziger:2021xvr}.

Consequently, there have been a number of recent efforts to reduce the
fraction $r_-$ of the negative weight contribution. One avenue that is
being explored is to optimise the event generation
itself~\cite{Frederix:2020trv,Gao:2020vdv,Bothmann:2020ywa,Gao:2020zvv,Danziger:2021xvr}. An
alternative approach is to remove negative weights from the
generated samples, while taking care that observables are not
affected~\cite{Andersen:2020sjs,Nachman:2020fff,Verheyen:2020bjw}. The
algorithm proposed in~\cite{Andersen:2020sjs} has been shown to be
very effective in a highly non-trivial example. However, there are two
caveats. First, the algorithm requires the choice of an auxiliary
distribution, despite being process-agnostic in nature. Second, while
the algorithm has been shown to produce sound results in practice, a
proof of the correctness has only been given for a specific set of
observables related to the chosen distribution.

In the following, we present a novel method, dubbed \emph{cell
resampling}, which eliminates negative event weights locally in phase
space, is independent of the underlying process, and preserves all
physical observables. Following the earlier
proposal~\cite{Andersen:2020sjs}, we insert an additional resampling
step before unweighting and detector simulation into the event generation chain
outlined above. This step only affects the event weights, leaving all
other properties untouched. Barring exact cancellations between
weights, the number of events remains the same as well. We describe
our method in section~\ref{sec:cellres}, and apply it to a calculation
of W boson production with two jets at next-to-leading order in
section~\ref{sec:application}. While we mostly focus on the
application to fixed-order QCD generation of weighted event samples,
the approach is largely independent of the details of the event
generation, and we comment on possible generalisations whenever we
have to make specific choices. We conclude with a summary and an outlook in
section~\ref{sec:conclusions}.

\section{Cell Resampling}
\label{sec:cellres}

Our main goal is to eliminate negative event weights without affecting
predictions for observables. We only consider observables that
can be both predicted faithfully by a given event generator and
measured in a real-world experiment involving detectors with a finite
resolution.

To measure an observable $\mathcal{O}$ we first select a phase-space
region $\mathcal{D}$ that is large enough to be resolved in a
real-world experiment, where the resolution is typically limited not
only by the detector, but also by the available statistics. The
measured value of the observable is then obtained by counting events
within this region. Obviously, this will always yield a non-negative
result.

On the theory side, we predict the value of the observable from a
Monte Carlo event sample with both positive and negative weights by
summing up the weights $w_i$ of the events $i$ contained in the
selected region $\mathcal{D}$:
\begin{equation}
  \label{eq:O_pred}
  \mathcal{O} = \sum_{i \in \mathcal{D}} w_i.
\end{equation}
Since this corresponds to a Monte Carlo estimate of the integrated
physical cross section over the selected phase space region, the
result has to be non-negative given enough statistics.

We now aim to modify the event weights in such a way that negative
weights are eliminated. At the same time, predictions for any
observables of the kind discussed above should be preserved. To this
end, we focus on a single event with negative weight, our \emph{seed},
and consider a small solid sphere in phase space that is centred
around said event and contains no other events. We call this sphere a
\emph{cell}. We then gradually increase the radius of the cell until
the sum over all weights of events contained inside the cell is
non-negative.

The key point is that with sufficiently many generated events, this
cell can be made arbitrarily small. In particular, it can be made so
small that it cannot be resolved anymore in a real-world
experiment. At this point, we may freely redistribute the weights of
the events within the cell without affecting any observables. One
possible choice is to replace all event weights by their absolute
value and rescale them such that the sum of weights is preserved. That
is, we perform the replacement
\begin{equation}
  \label{eq:resample}
  w_i \to \frac{\sum_{j\in\mathcal{C}} w_j}{\sum_{j\in\mathcal{C}} |w_j|} |w_i|
\end{equation}
for all weights $w_i$ of events $i$ inside the cell
$\mathcal{C}$. This accomplishes the original intent behind the
generation of negative-weight counterevents: any overestimation of the
cross section from positive-weight events is cancelled \emph{locally}
in phase space.

After applying the above steps successively to each negative-weight
event we are left with a physically equivalent sample of all
positive-weight events. While it is in principle possible to apply the
procedure to unweighted events, this would needlessly inflate cell
sizes and systematically discard all events inside a cell, since their
weights would sum up to zero. We therefore assume weighted input
events. If desired, efficient unweighting can be performed after the
elimination of negative weights.

Note that the method is completely agnostic to
any details of the event generation, including the underlying process,
and does not refer to any properties of specific observables. The only
prerequisite is sufficiently high statistics, so that single cells are
not resolved by any observable that can be measured in a real-world
experiment.

In practice, it may not be feasible to generate and resample enough
events to reach this goal. A particular concern is posed by cells
containing events in different histogram bins of some
distribution. Small enough cells close to bin boundaries can even turn
out to be beneficial by smoothing out statistical jitter due to bin
migration. However, as cells grow larger, the resampling will start to
smear out characteristic features, such as resonance peaks.

To ensure that smearing effects remain negligible compared to other
sources of uncertainty we can impose an upper limit on the cell
sizes. The cost we have to pay is that some cells may not contain
sufficient positive weight events to cancel the contribution from
events with negative weight. This implies that we no longer eliminate
all of the negative event weights. Still, since at least part of the
negative seed event weight is absorbed by surrounding events, a
subsequent unweighting will reduce the fraction of negative-weight
events compared to the original sample. The size of cells with
negative accumulated weight is related to the extent in phase space of
counter-terms, and as such is related to the usual problems with
bin-to-bin migration in higher-order calculations.

For the sake of a streamlined presentation, we will allow cells to
grow arbitrarily large for the remainder of this section. We will
come back to the discussion of a cell size limit when analysing the
practical performance of our method in section~\ref{sec:application}.

\subsection{Distances in Phase Space}
\label{sec:dist}

A central ingredient in cell resampling is the definition of a
distance function in phase space. While it may be possible to achieve
even better results with functions tailored to specific processes and
observables, our aim here is to define a universal distance measure.

There are two main requirements. First, let us consider events which
are close to each other according to our distance function. Such
events will very likely be part of the same cells and redistribute
weights among each other. It is therefore absolutely essential that
they have similar experimental signatures or only differ in properties
that are not predicted by the event generator. Otherwise it is
possible that only some of these events contribute to a given
observable and the total weight of this subset and therefore the
prediction for the observable is changed by the resampling. Second, it
is desirable that events with a large distance according to the chosen
function are easily distinguishable experimentally. In short, the
distance function should reflect both the experimental sensitivity and
the limitations of the event generator.

One consequence of these requirements is that the distance function
has to be infrared safe. Adding further soft particles to an event
should only change distances by a small amount. Specifically,
final-state particles with vanishing four-momentum are undetectable
and should not affect distances at all. Furthermore, the distance
function has to have limited sensitivity to collinear splittings as
well as soft radiation. In general, this can be achieved by basing the
distance function on suitable defined infrared-safe physics objects
instead of the elementary particles in the final state. The precise
definition of these physics objects may depend on the details of the
theory prediction. In the following, we consider only pure QCD
corrections and ensure infrared safety by considering final-state jets
instead of partons. It is natural to adopt the same jet clustering
that was used for the event generation.

Keeping the general requirements in mind, we now define a concrete distance
function for use with fixed-order Monte Carlo generators, which we use
in the following. Note that our choice is by no means unique and we do
not claim that our definition is optimal. A particularly promising
alternative is the ``energy mover's distance'' introduced
in~\cite{Komiske:2019fks} and generalised to include flavour
information in~\cite{Romao:2020ojy}. We leave a detailed comparison
between different distance functions to future work.

\subsubsection{Definition of the Distance Function}
\label{sec:d_def}

As a first step, we cluster the partons in each event into jets. We
then group outgoing particles into sets according to their types,
i.e. according to all discrete observable properties such as flavour
and charge. Here, we are using the word ``particle'' in a loose sense,
designating a jet --- or, more generally, any infrared-safe physics
object --- as a single particle. Particles which only differ by their
four-momenta (including possible differences in their invariant
masses) end up inside the same set. For the time being, we do not
differentiate between polarisations or various types of jets, but it
is straightforward to add such distinctions.

The distance $d$ between an event $e$ with $T$ particle type sets
$\mathcal{S} = \{s_1, s_2,\dots, s_T\}$ and an event $e'$ with sets
$\mathcal{S'} = \{s_1', s_2',\dots, s_T'\}$ is then given by the sum
of the distances between matching sets, i.e.
\begin{equation}
  \label{eq:dist_E}
  d(e, e') = \sum_{t=1}^T d(s_t,s_t')\,,
\end{equation}
where $s_t, s_t'$ contain all particles of type $t$ that occur in the
respective event. If there are no particles of type $t$ in a given
event, then the corresponding set is empty. This ensures that all
events in the sample have the same number of particle type sets.

Next, we have to define the distance between a set $s_t$, containing
$P$ particles of type $t$ with four-momenta $p_1,\dots,p_P$ and a set $s_t'$ with
$Q$ particles of the same type with four-momenta
$q_1,\dots,q_Q$. Without loss of generality we can assume $Q\leq
P$. To facilitate the following steps we first add $P-Q$ particles
with vanishing momenta to the set $Q$, i.e. we define the momenta
\begin{equation}
  \label{eq:q_aux}
  q_{Q+1} = \dots = q_P = 0\,.
\end{equation}
Naturally, the distance between $s_t$ and $s_t'$ must not depend on the
labelling of the momenta. We therefore consider all permutations of the
momenta $q_1,\dots,q_P$ and sum the pairwise distances to the momenta
$p_1,\dots,p_P$ for each permutation. The minimum defines the distance
between $s$ and $s'$:
\begin{equation}
  \label{eq:d_set}
  d(s_t,s_t') = \min_{\sigma \in S_P} \sum_{i=1}^P d(p_i, q_{\sigma(i)})\,.
\end{equation}
Here, $S_P$ denotes the symmetric group, i.e. the group of all permutations
of $P$ elements.

While this distance function is obviously invariant under relabelling
and when adding vanishing momenta, its calculation quickly becomes
prohibitively expensive when the number of momenta $P$ in the sets
grows large. More precisely, the computational cost scales as $\mathcal{O}(P!)
= \mathcal{O}\Bigl(P^{P+\frac{1}{2}}\Bigr)$. For events obtained through fixed-order
calculations, $P$ is typically small enough and the poor scaling is
not a concern. However, after parton showering, large values of $P
\approx 10$ can be reached. In such cases, we use an
alternative distance function $\tilde{d}$ which is easier to
compute. We first define a unique labelling by ordering the momenta
according to their norm
\begin{equation}
  \label{eq:norm}
  \| p_i \| \equiv d(p_i, 0)\,,\qquad\| q_i \| \equiv d(q_i, 0)\,.
\end{equation}
We then search for the
nearest neighbour of $p_1$ in $Q$, i.e. the momentum
$q_{\text{NN}(1)} \in Q$ that minimises the distance to $p_1$:
\begin{equation}
  \label{eq:nearest_neighbour_dist}
  d(p_1, q_{\text{NN}(1)}) \leq d(p_1, q_i) \qquad \forall q_i \in s_t'\,.
\end{equation}
In the next step, we find the nearest neighbour $q_{\text{NN}(2)}$ of
$p_2$, \emph{excluding} the momentum $q_{\text{NN}(1)}$ and add
$d(p_2, q_{\text{NN}(2)})$ to the total distance. We iteratively define
further exclusive nearest neighbours $\text{NN}(j)$ by removing previous ones,
\begin{equation}
  \label{eq:nearest_neighbour_set}
  s'_{t,\overline{\jmath}} = s_t' \backslash \{q_{\text{NN(1)}},\dots,q_{\text{NN(j-1)}}\}\,,
\end{equation}
and finding the element $q_{\text{NN}(j)}$ such that
\begin{equation}
  \label{eq:nearest_neighbour_j}
  d(p_j, q_{\text{NN}(j)}) \leq d(p_j, q_i) \qquad \forall q_i \in s_{t,\overline{\jmath}}'\,.
\end{equation}
Finally, we add up all these distances. To arrive at a symmetric
distance function, we compare to the total distance obtained by
exchanging $s_t$ and $s_t'$. The smaller of the two distances
defines $\tilde{d}(s_t, s_t')$:
\begin{equation}
  \label{eq:d_tilde}
  \tilde{d}(s_t,s_t') = \min\left(\sum_{i=1}^P d(p_i, q_{\text{NN(i)}}),\ \sum_{i=1}^P d(q_i, p_{\text{NN(i)}})\right)\,.
\end{equation}
Since this minimises over only two out of all possible sets of
pairings between momenta $p_i$ and $q_j$, $\tilde{d}(s_t,s_t')$ scales as $\mathcal{O}(P^2)$ instead of
$\mathcal{O}\Bigl(P^{P+\frac{1}{2}}\Bigr)$. For the same reason, it is bounded from below by
$d(s_t,s_t')$. However, for nearby events we expect that the exclusive
nearest neighbours $\text{NN}(i)$ defined above coincide with the
\emph{actual} nearest neighbours. In this case the distance functions
$\tilde{d}$ and $d$ will yield the same result. Let us finally remark
that $\tilde{d}$ also remains unchanged when adding vanishing momenta
to the sets $s_t$ and $s_t'$.

The last missing ingredient is the distance $d(p,q)$ between two momenta $p$
and $q$. One can in principle choose different distance functions for
different particles types (e.g.~to include only transverse momentum for
neutrinos), but for simplicity we will in this study apply the same distance
for all types. Note that the Minkowski distance
$\sqrt{(p-q)_\mu (p-q)^\mu}$ is \emph{not} a suitable distance. Such a
distance would be insensitive to lightlike differences between momenta, which
does not match the reality of experimental measurements. Since, for a given
particle mass, only three momentum components are independent, our distance
measure is based on the difference in \emph{spatial} momentum. An\-ti\-ci\-pa\-ting
the fact that experimental measurements are more sensitive to the
perpendicular momentum components we further add a rescaled difference in
transverse momenta:
\begin{equation}
  \label{eq:norm_eucl}
  d(p,q) = \sqrt{\sum_{i=1}^3 (p_i - q_i)^2 + \tau^2 (p_\perp - q_\perp)^2}\,.
\end{equation}
$\tau$ is a tunable parameter.

We emphasise again that the presented distance function is purely
intended for use with fixed-order QCD, and we are only concerned about
sensitivity to those observables that can be predicted
faithfully. More sophisticated theory predictions, like those obtained
from parton showers, will require further refinements. For example,
sensitivity to jet substructure could be introduced through a
specialised metric for the distance between two jets in place of
equation~\eqref{eq:norm_eucl}. We will leave such explorations to
future work.

\subsubsection{Alternative Norms}
\label{sec:alt_norms}

The choice of a momentum distance function is not unique and
one could alternatively consider other coordinate systems, such as
light-cone coordinates, or general p-norms
\begin{equation}
  \label{eq:norm_p}
  d_p(p,q) = \left(\sum_{i=1}^3 (p_i - q_i)^p\right)^{\frac{1}{p}}.
\end{equation}
One interesting possibility yet to be explored is to take inspiration
from jet distance measures and employ a distance in rapidity or
azimuthal angle, which has to be combined with a transverse momentum
distance in a meaningful way. We leave such explorations to further
study.

It is, in any case, crucial to not violate the triangle inequality
\begin{equation}
  \label{eq:tri}
  d(p,q_1 + q_2) \leq d(p,q_1) + d(p,q_2).
\end{equation}
To illustrate this point, let us consider the \emph{square} of the spatial norm
\begin{equation}
  \label{eq:norm_eucl_sq}
  d^2(p,q) = \sum_{i=1}^3 (p_i - q_i)^2,
\end{equation}
a set $s$ with one particle momentum $p_1$, and a set $s'$ with one
particle momentum $q_1$. For simplicity, let us assume one-dimensional
momenta with $p_1=2$ and $q_1=-2$ in arbitrary units. Since there is only
one possible pairing of momenta, the event distance is $d_2(s,s') = d(p_1,
q_1) = 4$ using the spatial $p=2$ norm and $d^2(s,s') = d^2(p_1,q_1) = 16$ when
using the square.

\begin{figure}[htb]
  \centering
  \includegraphics[width=0.45\linewidth]{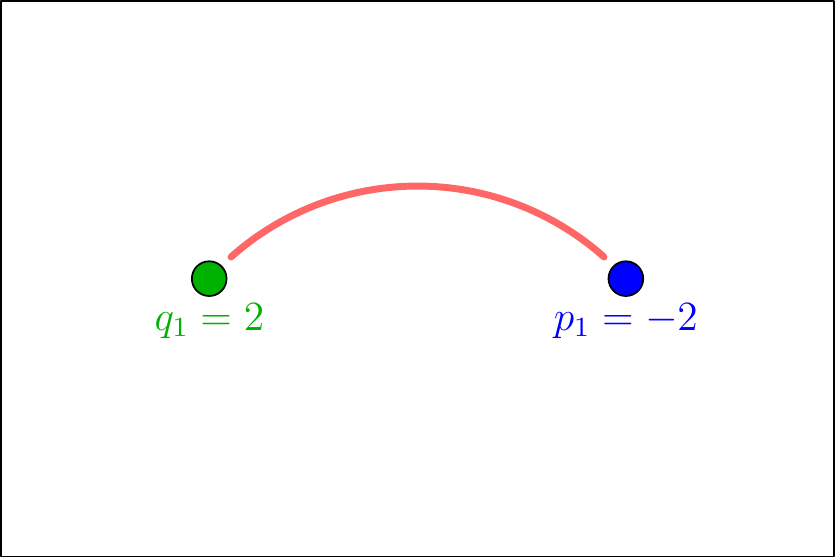}~\includegraphics[width=0.45\linewidth]{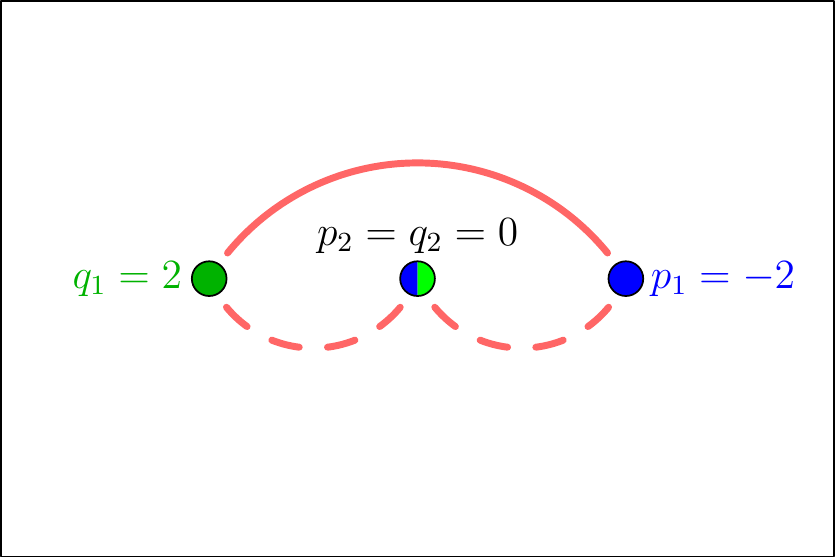}
  \caption{Distances between two sets $s, s'$, each of which contains one particle (left panel) or two particles (right panel). Blue dots represent particles in set $s$, green dots stand for particles in set $s'$. The possible pairings used in the distance calculation are indicated by lines. The square norm $d^2$ erroneously results in a shorter distance for the pairing indicated by the dashed line.}
  \label{fig:triangle}
\end{figure}

We now add one particle with vanishing momentum to each $s$ and $s'$,
reminding ourselves that this change is not experimentally detectable
and therefore must not affect the distance between the sets. This
scenario is shown in figure~\ref{fig:triangle}. According to
equation~\eqref{eq:d_set} the set distance is now given by
\begin{equation}
  \label{eq:dist_example_eucl} d(s, s') = \min\Bigl[d(p_1, q_1) +
d(p_2, q_2), d(p_1, q_2) + d(p_2, q_1)\Bigr],\qquad p_2 = q_2 = 0\,,
\end{equation}
with $d(p_2, q_2)=0$ for either choice. For the spatial norm both
permutations yield the same result and $d_2(s,s') = 4$ as
before. However, if we use the square $d^2$ instead, we find that the
triangle inequality is violated and $d^2(s, s') = 8$ instead of 16.

\subsubsection{Example for an Event Distance}
\label{sec:dist_example}

To illustrate one possible distance measure, let us consider a simple yet
non-trivial example, where we calculate the distance between two
events $e$ and $e'$. $e$ has two outgoing photons and a jet, and $e'$
two photons and no jets. The distance between the two events is
therefore
\begin{equation}
  \label{eq:d_example_events}
  d(e, e') = d(s_\gamma, s_\gamma') + d(s_j, s_j').
\end{equation}
The jet set $s_j$ for the event $e$ contains a single jet with
momentum $p_j$, whereas the corresponding set $s_j'$ for $e'$ is
empty. Likewise, $s_\gamma$ contains the photons in $e$ with momenta
$p_1, p_2$ and $s_\gamma'$ the photons in $e'$ with momenta $p_1',
p_2'$. The distance between these two sets is
\begin{equation}
  \label{eq:d_photons_example}
  d(s_\gamma, s_\gamma') = \min\Bigl[d(p_1,p_1') + d(p_2, p_2'), d(p_1,p_2') + d(p_2, p_1')\Bigr],
\end{equation}
where $d(p,q) = \sqrt{\sum_{i=1}^3 (p_i-q_i)^2 +
\tau^2(p_\perp-q_\perp)^2 }$ is the distance introduced in
equation~\eqref{eq:norm_eucl}.

To calculate the distance $d(s_j, s_j')$ between the jet sets, we
first add an auxiliary ``jet'' with vanishing momentum $p_j' = 0$ to
$s_j'$, so that both sets each contain a single jet. The set distance
is then
\begin{equation}
  \label{eq:d_jets_example}
  d(s_j, s_j') = d(p_j, p_j') = \sqrt{\sum_{i=1}^3 p_{ji}^2 + \tau^2 p_{j\perp}^2},
\end{equation}
where $p_j$ denotes the momentum of the physical jet.

\subsubsection{Distances in high-dimensional spaces}
\label{sec:dimensions}

When considering a fixed number of generated events in phase spaces of
increasing dimension, the typical distance between two events, and
therefore the characteristic cell size, will grow very
quickly. Conversely, when trying to achieve a given cell size, the
required number of events will become prohibitively large very soon.
We might be worried that for processes with many final-state particles
cells unavoidably become so large that they can be resolved
experimentally. This is not the case. The reason is that experimental
resolution is limited by statistics in addition to detector
limitations. In fact, the sizes of resolved regions of phase space
will increase in the same way as the cell sizes with the number of
dimensions.

It is instructive to consider how the final cell size in a
high-dimensional phase space translates to differences in real-world
observables, which are typically low-dimensional distributions. To
obtain some quantitative insight, we can model a cell as an $n$-ball
with radius $R$ in a $n$-dimensional Euclidean space. For a
sufficiently small radius, events within the cell will be
approximately uniformly distributed. The exceptions are the cell seed
at the centre and the last event added at a distance $R$ to the
cell. It is well known that for large $n$, events tend to be close to
the surface, and the mean distance of an event to the seed is indeed given by
\begin{equation}
  \label{eq:d_mean}
  \bar{d} = \frac{1}{V_n} \int dV_n\ r = \frac{n}{n+1}R,
\end{equation}
where $V_n$ is the volume of the $n$-ball. However, when predicting
one-dimensional distributions, we project the events onto a line,
which we can take as the $n$-th coordinate axis. For the mean distance in this direction we obtain
\begin{equation}
  \label{eq:dn_mean}
  \bar{d}_n = \frac{1}{V_n} \int dV_n\ r_n = \frac{\Gamma(\frac{n}{2}+1)}{\Gamma(\frac{1}{2})\Gamma(\frac{n+3}{2})}R = \mathcal{O}\left(\frac{1}{\sqrt{n}}\right).
\end{equation}
The difference in one-dimensional distributions is much smaller than
the cell size, especially in high-dimensional phase spaces.

\subsection{Nearest-Neighbour Search and Locality-Sensitive Hashing}
\label{sec:lsh}

\begin{figure}[htb]
  \centering
  \includegraphics{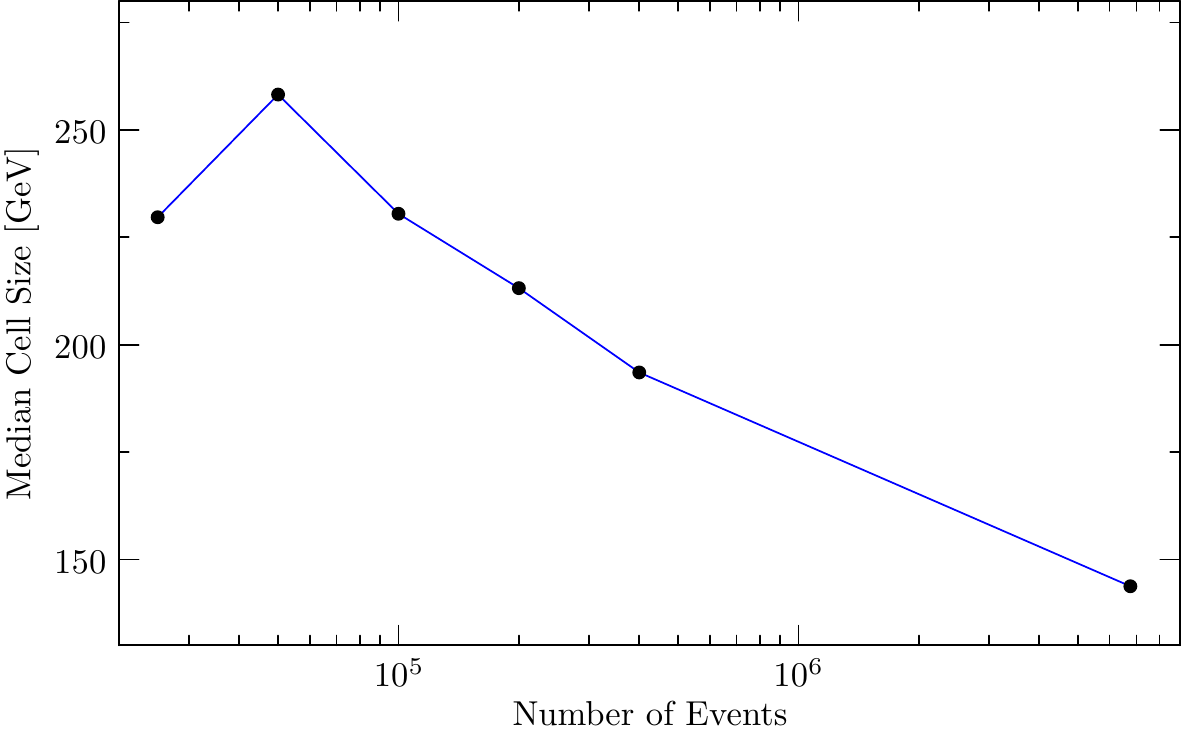}
  \caption{Median cell radius as a function of sample size according
to the distance measure defined in section~\ref{sec:d_def} with
$\tau=0$. See section~\ref{sec:application} for details on the event
generation.}
  \label{fig:size_scaling}
\end{figure}

A very appealing feature of cell resampling is that the cell radii
automatically become smaller with increasing  number of
events. Anticipating the discussion of the practical application
with the distance measure in section~\ref{sec:application}, we indeed find a
significant reduction in cell size, as shown in
figure~\ref{fig:size_scaling}. Due to large fluctuations in the
weights of the cell seeds, and correspondingly in the cell sizes, we
consider the median of the cell radii instead of the arithmetic mean.

From this point of view, cell resampling should be applied to event
samples that are as large as possible. However, the main motivation
behind the resampling is to reduce the computational cost of later
steps. This implies that the CPU time needed for resampling should be
much smaller. On the one side, various event generation steps are
usually performed on an event-by-event basis, and the asymptotic
generation time scales linearly with the number $n$ of events. On the
other side, the number of negative-weight events and therefore the
number of cells are also proportional to $n$. To achieve the same
asymptotic $\mathcal{O}(n)$ runtime scaling, the time to construct a single
cell has to be constant, i.e. independent of the sample size. However,
while obviously asymptotically inferior, overall $\mathcal{O}(n \log^k
n)$ scaling for some integer $k$ will presumably be sufficient for
real-world sample sizes.

To reiterate, a cell is constructed as follows. We choose a seed,
i.e. one of the remaining negative-weight events. As long as the sum
of weights inside the cell is negative, we subsequently find the
closest event to the seed outside the cell according to our distance
function and add it to the cell. Finally, we reweight all events
inside the cell.

The selection of a negative-weight event can be achieved in constant
time with the help of a pregenerated array containing these events,
which can be created in $\mathcal{O}(n)$ steps. If we want to
construct cells in a specific order, an additional $\mathcal{O}(n \log
n)$ sorting step is required. The impact of different seed selection
strategies is discussed in appendix~\ref{sec:seed_strategies}.

The biggest challenge is to find nearby events in constant or, at
worst, logarithmic time. A naive nearest-neighbour search requires
$\mathcal{O}(n)$ operations for each cell, leading to an overall
$\mathcal{O}(n^2)$ scaling. Nearest-neighbour search is a well-studied
problem in computer science. In modern particle physics, one of the
most prominent applications is jet
clustering~\cite{Cacciari:2005hq}. However, jet clustering requires
searches in two dimensions and the algorithms used there are not
suited for the present high-dimensional problem.  For approximate
searches in high-dimensional spaces, \emph{locality-sensitive hashing}
(LSH)~\cite{Indyk1998,Leskovec:2020} has been found to be highly
efficient. In the following we give a short outline while describing
one way in which LSH can be used for finding events that are close to
a given cell.

First, we relax the condition that cells have to be kept spherical by
subsequently adding nearest neighbours. We will contend ourselves with
cells that are small in each phase-space direction and, most
importantly, shrink with increasing statistics. This can be achieved
by an approximate nearest-neighbour search.

In order to being able to apply standard techniques, we map the events
onto points in a high-dimensional Euclidean space. A crucial
observation is that this map does not have to preserve distances,
neither in the absolute nor in the relative sense. It is sufficient
that events that are nearby in phase space are mapped onto points that
are nearby in Euclidean space with high probability. To this end, we
recall that, in each event $e$, we group particles into sets
$s_{e,t}$ according to their type $t$. To each of these sets, we now
add particles with vanishing four-momentum until for each particle
type all sets in all events have the same number of
elements,\footnote{Sets corresponding to different particle types may
still have different cardinalities.} i.e.
\begin{equation}
  \label{eq:set_size}
  |s_{e,t}| = |s_{e',t}|\qquad \text{for all events }e,e'\text{ and types }t\,.
\end{equation}
We then interpret each set $s_{e,t}$ as a tuple $\tau_{e,t} =
\big(p_{e,t,1},\dots,p_{e,t,P_t}\big)$ of $P_t$ four-momenta, which we
sort lexicographically according to the momentum components. Finally,
for the event $e$ the coordinates of the resulting point $V_e$ in
Euclidean space correspond to the spatial momentum components in order. The
first three components $i=1,2,3$ are obtained from the first spatial momentum
in the tuple for the first particle type, i.e. $(V_{e})_i =
(p_{e,1,1})_i$. The next three components correspond to the second
momentum in the tuple for the first particle type, and after reaching
the last momentum in a tuple we proceed to the tuple for the next
particle type. The total dimension of the Euclidean space is therefore
\begin{equation}
  \label{eq:dim_eucl}
  D = 3 \sum_{t=1}^T |s_{e,t}|\,,
\end{equation}
which is independent of the chosen event $e$. Events $e$ and $e'$
which are nearby in phase space will most likely have a similar number
of outgoing particles of each type with similar momenta. Therefore,
the coordinates of $V_e$ and $V_{e'}$ will be similar and the points
will be close to each other.

We have now recast the problem to finding approximate nearest
neighbours between the points $V_e$ in a $D$-dimensional Euclidean
space. To facilitate this task, we consider \emph{random projections}
onto lower dimensional sub spaces. This is a well-established
technique motivated by the Johnson-Lindenstrauss
Lemma~\cite{JohnsonLindenstrauss}, which states that it is always
possible to find projections that nearly preserve the distances
between the points. Specifically, we define projections $\Pi_h$ onto
hyperplanes $h=1,\dots,H$, each fixed by a randomly chosen unit normal
vector as discussed in~\cite{Muller:1959,Marsaglia:1972}. $H$ should be chosen in such
a way that it grows logarithmically with the number of points,
i.e. the number of events. We then perform a second projection onto
the $x$ axis, i.e. we take the first coordinate of each projected
vector:
\begin{equation}
  \label{eq:proj}
  V_e \mapsto c_{e,h} = \left(\Pi_h V_e\right)_1.
\end{equation}
For each $h=1,\dots,H$, we sort the coordinates $c_{e,h}$ obtained
from all events $e$ in the sample and divide them into \emph{buckets}
of size $B$. $B$ may increase at worst logarithmically with the number
of events, otherwise the described algorithm for cell creation would
no longer fulfil our time complexity constraint, as shown later. Note
that events $e, e'$ with nearby coordinates $c_{e,h}, c_{e',h}$ will
be inside the same bucket with high probability.

This effectively defines $H$ \emph{hash functions} $f_h$, each of
which maps a given event onto a single integer number, namely the
index of the bucket associated with the coordinate $c_{e,h}$:
\begin{equation}
  \label{eq:hash_fun}
f_h:\qquad e \mapsto \bucket\left[c_{e,h}\right] = \bucket\left[\left(\Pi_h V_e\right)_1\right]\,.
\end{equation}
These hash functions are \emph{locality sensitive}. Nearby events are
likely to be mapped onto the same number.

We use the hash functions to create $H$ \emph{hash tables}, where we
store the values of $f_h(e)$ for all events $e$. When creating a cell,
we look up the hashed values $f_h(e)$ for the seed event $e$. We then
consider all events that lie in the same bucket as $e$ in one of the
$H$ projected coordinates, i.e. at most $HB$ events. In practice, the
number will be much lower, as nearby events will share more than one
bucket with $e$. Starting from the events sharing the most buckets
with $e$, we identify nearest neighbour candidates.\footnote{In
principle, one could consider all events sharing buckets with $e$
without violating the complexity constraints. However, in practice we
only need a small fraction of these events to create the cell.} We
select candidates with weights $w_i$, until we fulfil
\begin{equation}
  \label{eq:cand_weight}
  \sum_i w_i \geq -a w_e\,,
\end{equation}
where $w_e$ is the (negative) seed weight and $a$ an arbitrary constant
with $a \geq 1$. In our implementation, we choose $a=2$.

For each of the nearest neighbour candidate, we then compute the
actual distance to the seed $e$ using the metric defined in
section~\ref{sec:dist}. We proceed to add nearest neighbours
according to the actual distance to the cell, as we did when using
naive nearest-neighbour search, until the sum of weights inside the
cell is no longer negative.

Let us now analyse the asymptotic time complexity of the LSH-based
algorithm. To create the hash table, we have to compute $H$ projected
coordinates for each of the $n$ events, where $H = \mathcal{O}(\log
n)$. Once the $H$ hyperplanes are fixed, each coordinate can be
computed independently in constant time. In total we compute
$\mathcal{O}(H n) = \mathcal{O}(n \log n)$ coordinates. For each $h$,
we have to sort the coordinates of $n$ events, which gives a time
complexity of $\mathcal{O}(H n \log n) = \mathcal{O}(n \log^2 n)$. The
partitioning into buckets is again linear in the number of events, and
therefore asymptotically negligible compared to the sorting.

Looking up the values of the hash functions for a single seed event
requires $\mathcal{O}(H) = \mathcal{O}(\log n)$ operations. To select
the candidates we have to probe all events in $H$ buckets of size $B$,
with time complexity $\mathcal{O}(HB) = \mathcal{O}(\log^2 n)$. Since
the number of events needed to compensate the seed weight does not
increase with $n$, only a constant number of distance functions have
to be calculated after selecting candidates. In total, the number of
cells grows linearly with $n$ and the creation of each cell requires
$\mathcal{O}(\log^2 n)$ constant-time steps, which results in an
overall time complexity of $\mathcal{O}(n \log^2 n)$.

\begin{figure}[htb]
  \centering
  \includegraphics[width=0.45\linewidth]{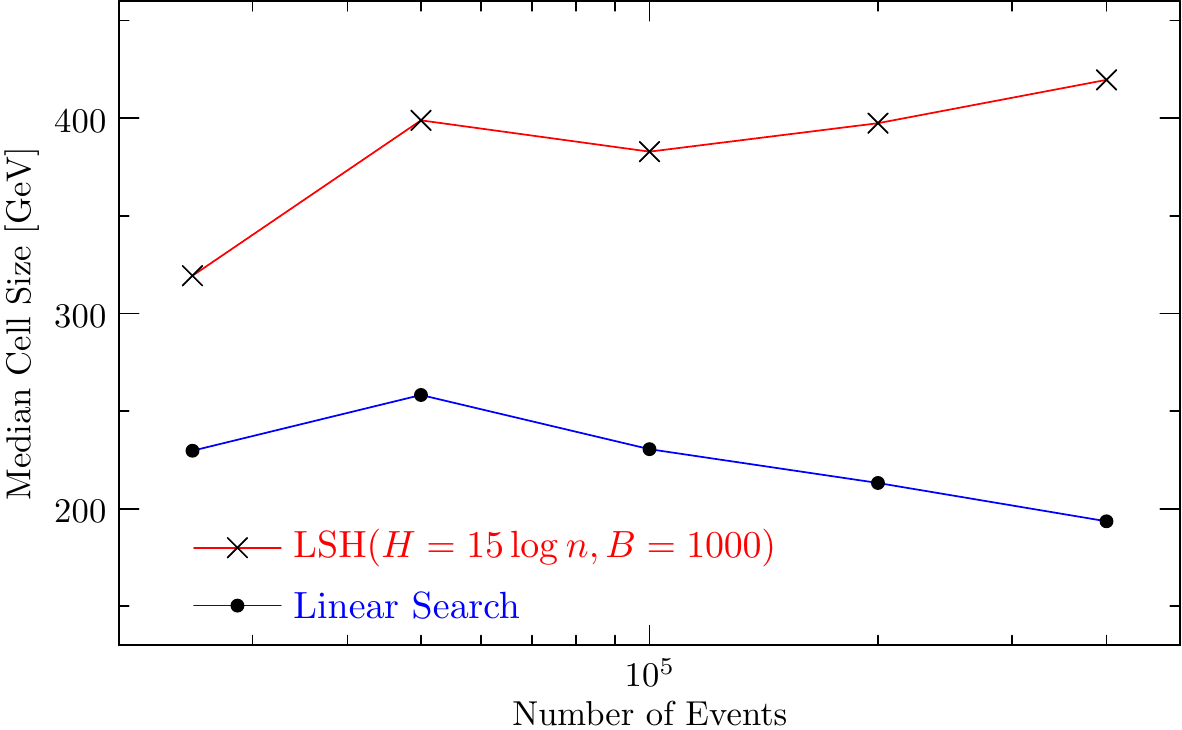}\qquad\includegraphics[width=0.45\linewidth]{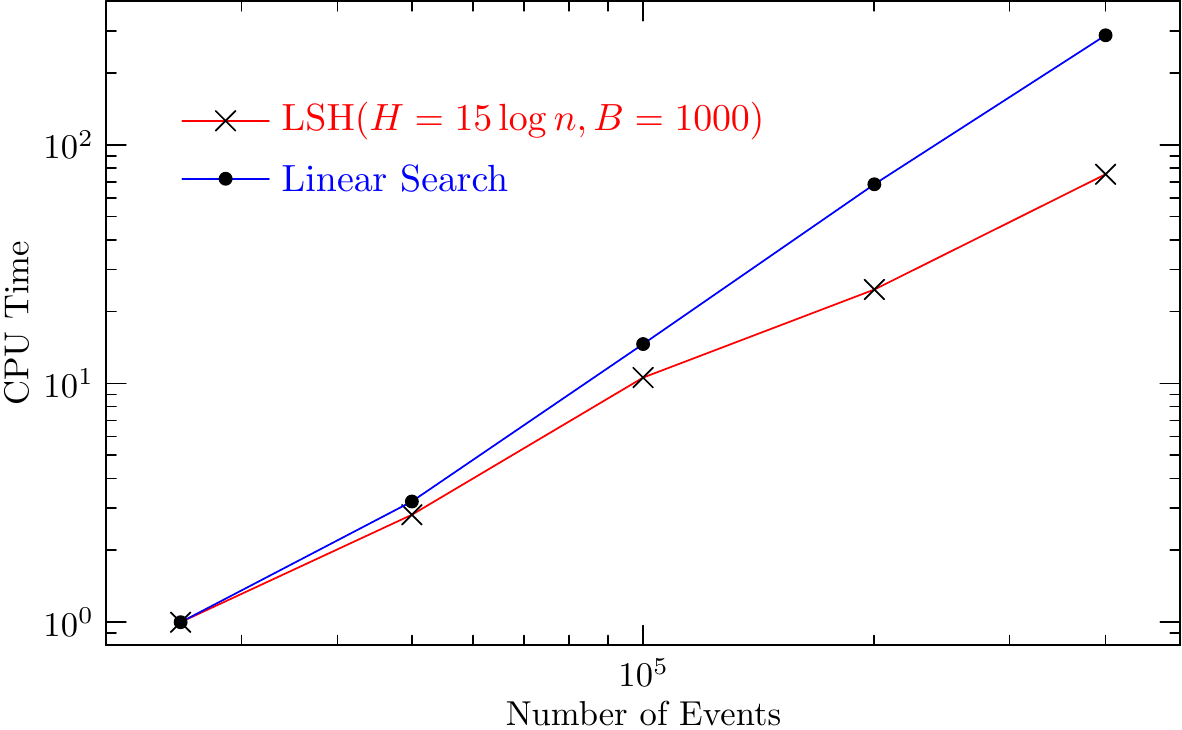}
  \caption{Comparison between linear nearest-neighbour search and the
LSH-based approach with $H=15\log n$ hash functions and buckets containing
$B=1000$ events. The left panel shows the median cell size as a
function of the number of input events. The right panel illustrates
the scaling of the computing time with the sample size. The respective
timings are normalised to the first data point in order to facilitate
comparison and eliminate constant factors resulting from differences
in the implementation.}
  \label{fig:cmp_lsh}
\end{figure}

Finally, we consider the practical performance of our implementation
for various sample sizes. Figure~\ref{fig:cmp_lsh} compares the median cell size,
measured by the largest distance between the seed and any other event
inside the cell, and scaling of the computing time to the naive linear
nearest-neighbour search. Heuristically, we have chosen
 $H=15\log n$ hash functions and a fixed bucket size of $B=1000$. We
find that the required computing time indeed scales much better with
increased sample size. However, the cell size does not seem to
decrease significantly with larger statistics. We conclude that
further improvements are needed to render the LSH-based approach
competitive compared to the linear nearest-neighbour search.

\subsection{Relation to Positive Resampling}
\label{sec:pos_res}

Let us comment on the relation between the current approach and the
positive resampler suggested in~\cite{Andersen:2020sjs}. There, events
are collected in histogram bins and positive resampling,
i.e. replacing event weights by rescaled absolute values, is performed
on each bin separately.

There is a clear parallel to the LSH-based method presented in
section~\ref{sec:lsh}. Where the LSH includes random projections of
events onto coordinates, positive resampling projects events onto one
or more observables defined by the chosen histogram, such as particle
transverse momenta. The histogram bins then correspond to the buckets
in the locality-sensitive hash tables. Following the same philosophy
behind the random projection method, we expect events in the same
histogram bins to be also close in other phase-space directions with
high probability.

\begin{figure}[htb]
  \centering
  \includegraphics{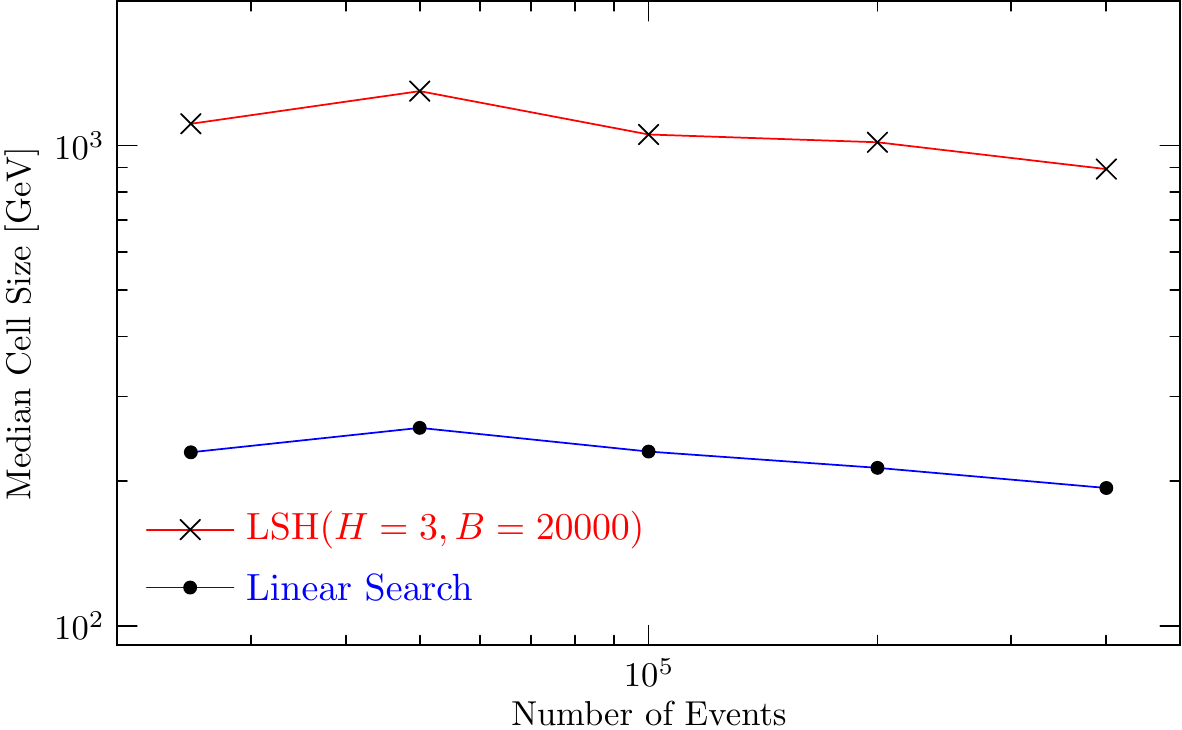}
  \caption{Comparison of the median cell size between linear
nearest-neighbour search and the LSH-based approach with three hash
functions.}
  \label{fig:cmp_lsh3}
\end{figure}

Positive resampling as presented in~\cite{Andersen:2020sjs} is
therefore similar to cell resampling with LHS-based nearest-neighbour
search with a small number of $H \leq 3$ hash functions. An important
difference is the definition of a cell. The equivalent for the
positive resampler would be an entire bin, i.e a complete hash
bucket. For this reason, it is difficult to perform a rigorous direct
comparison between the two approaches. However, we can get an idea by
comparing cell resampling with linear nearest-neighbour search and
LSH-based search with small $H$. In figure~\ref{fig:cmp_lsh3} we compare the
median cell sizes obtained in both approaches for different sample
sizes. We have increased the bucket size to $B=20000$ to ensure that
sufficiently many nearest-neighbour candidates are found.

\section{Cell Resampling for the Production of a  W Boson with two Jets}
\label{sec:application}

We now demonstrate the proposed method through a highly non-trivial
application. We consider the production of a leptonically decaying
W$^-$ boson together with at least two jets in proton-proton
collisions at 7 TeV centre-of-mass energy calculated at
next-to-leading order. Weighted events are generated using
Sherpa~\cite{Bothmann:2019yzt} with
OpenLoops~\cite{Buccioni:2019sur}. Generation parameters are shown in
table~\ref{tab:gen}.
\begin{table}[htb]
  \centering
  \begin{tabular}{l@{\hspace{2cm}}ll}
    \toprule
    \# events                  & \multicolumn{1}{l}{$6.1\times 10^7$} & (low-statistics input sample)     \\
                               & \multicolumn{1}{l}{$1.5\times 10^{10}$}   & (high-statistics reference sample)    \\
    $\mu_r, \mu_f$             & \multicolumn{2}{l}{$\frac{H_T}{2}$}                                \\
    $\sqrt{s}$                 & \multicolumn{2}{l}{7\,TeV}                                         \\
    PDF                        & \multicolumn{2}{l}{NNPDF 3.1~\cite{NNPDF:2017mvq,Buckley:2014ana}} \\
    Jet definition             & \multicolumn{2}{l}{anti-$k_t$~\cite{Cacciari:2008gp}}              \\
                               & \multicolumn{2}{l}{$R = 0.4$}                                      \\
                               & \multicolumn{2}{l}{$p_\perp > 30\,$GeV}                            \\
                               & \multicolumn{2}{l}{$|\eta| < 4.4$}                                 \\
    $m_W$                      & \multicolumn{2}{l}{80.385\,GeV}                                    \\
    $\Gamma_W$                 & \multicolumn{2}{l}{2.085\,GeV}                                     \\
    $1/\alpha_{\text{QED}}(0)$ & \multicolumn{2}{l}{132.232}                                        \\
    \bottomrule
  \end{tabular}
  \caption{Parameters used for event generation.}
  \label{tab:gen}
\end{table}

The 2-parton samples receive contributions from Born, virtual, and
subtraction terms, and the 3-parton samples from subtraction and real
emission. The sum should be positive, provided reasonable choices for
the renormalisation and factorisation scales have been applied. We
choose $\mu_f=\mu_r=H_T/2$. The subtraction terms can extend far in
phase space, which can cause non-local bin-to-bin migration in an
analysis of distributions, and issues for the current resampling based
on phase space buckets. This issue can be reduced by applying the
modified subtractions~\cite{Nagy:2001fj,Nagy:2003tz} to restrict their
size. We have not done so in the current study, partly to illustrate
the performance of the resampler in a ``worst case'' scenario.

\begin{figure}[htbp]
  \centering
  \includegraphics{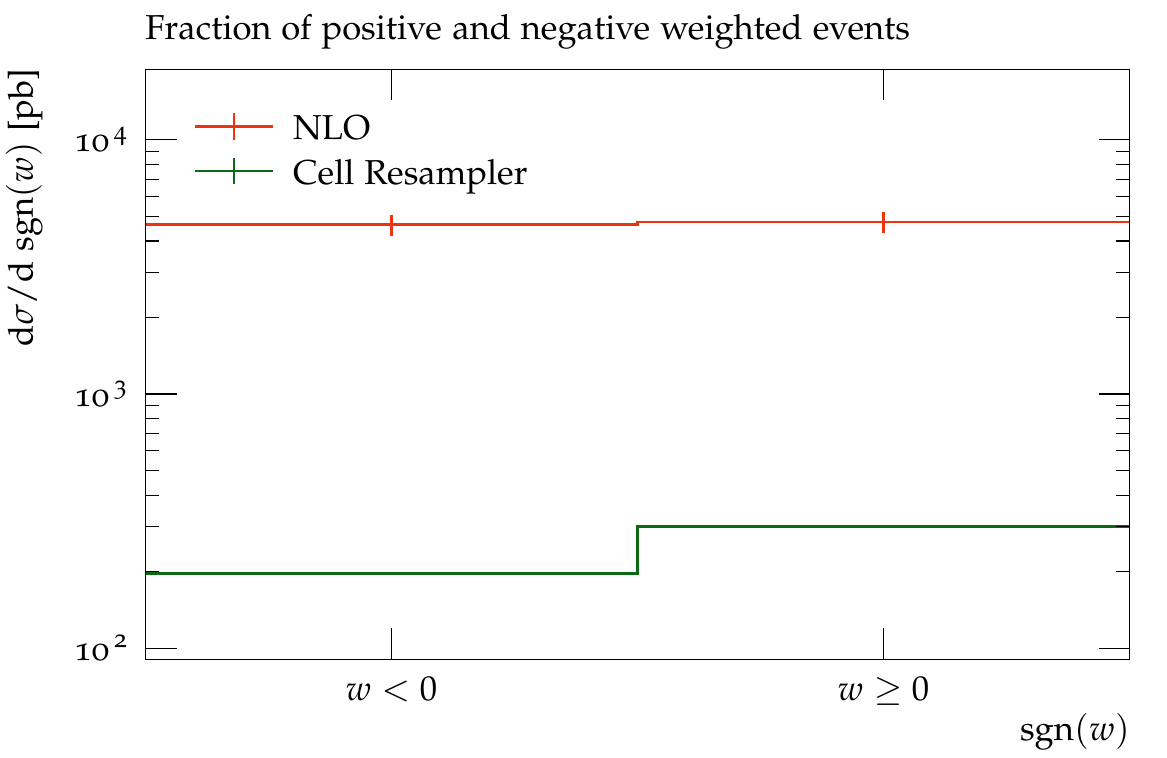}
  \caption{Contributions of negative- and positive-weight events to the cross section according to the \texttt{MC\_XS} \textsc{Rivet} analysis.}
  \label{fig:neg_contr}
\end{figure}

To assess the performance of our method, we use $6.1\times 10^7$
events as input and compare both this original NLO sample and the
output of the cell resampler to a reference prediction obtained from a
high-statistics sample with $1.5\times 10^{10}$ events. For the cell
resampling, we choose the distance measure introduced in
section~\ref{sec:d_def}, and determine the contribution from
transverse momentum differences by setting the parameter $\tau$ to 10,
c.f. equation~\eqref{eq:norm_eucl}. Since the computing time required
for cell resampling grows quadratically with the number of events, we
split the input sample into nine samples of equal size and apply
the resampling separately to each of them. At the same time, we impose
an upper limit for the cell radius of 100\,GeV in order to achieve
cell sizes commensurate with the combined sample size. While this
limit prevents us from eliminating \emph{all} negative event weights,
it still reduces their overall contribution by more than an order of
magnitude. This is illustrated in figure~\ref{fig:neg_contr}.

The NLO calculation achieves a $\sim 10^2\,$pb cross section as cancellation
between terms of size $\sim 5\times 10^3\,$pb, whereas the cell resampler
obtains the same cross section as a difference between terms of
$\sim 3\times 10^2\,$pb and $\sim 2\times 10^2\,$pb. If we were to reweight the events
to weight $\pm 1$, the original NLO sample would result in almost $20$
times as many events compared to unweighting the resampled
events. This corresponds to a Monte Carlo estimate of the uncertainty
which is larger by a factor of approximately $\sim
\sqrt{20}$. Conversely, to achieve the same estimated uncertainty as reached
after resampling, a pure NLO sample would require the generation of
almost 20 times as many events.

The events are parsed through the standard
\textsc{Rivet}~\cite{Bierlich:2019rhm} analyses \texttt{MC\_XS} and
\texttt{MC\_WJETS}, plus a relevant ATLAS analysis~\cite{ATLAS:2014fjg} in order to
demonstrate the performance for calculations relevant for experimental
measurements. As the point of this study is not to compare with experimental
data, but to improve on the stability of the predictions, we will for
simplicity use the calculation of just $pp\to e^- \bar \nu_e j j$ and not the
positron channel. In addition, we also extract the rapidity
distribution of the W boson using a custom analysis based on the
\textsc{Rivet} framework.

\begin{figure}[htbp]
  \centering
  \begin{tabular}{cc}
    \includegraphics[width=0.44\linewidth]{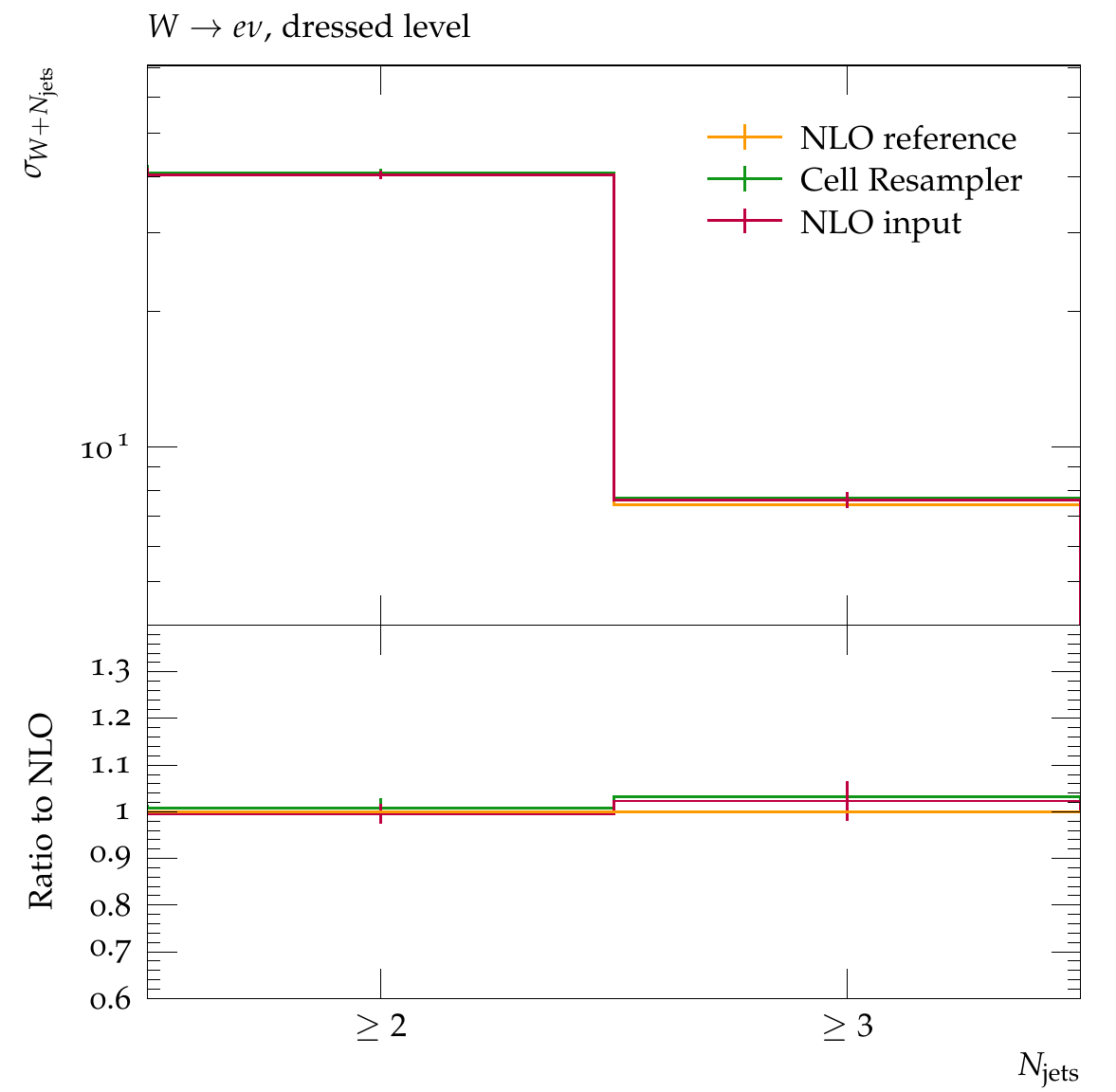}&
    \includegraphics[width=0.44\linewidth]{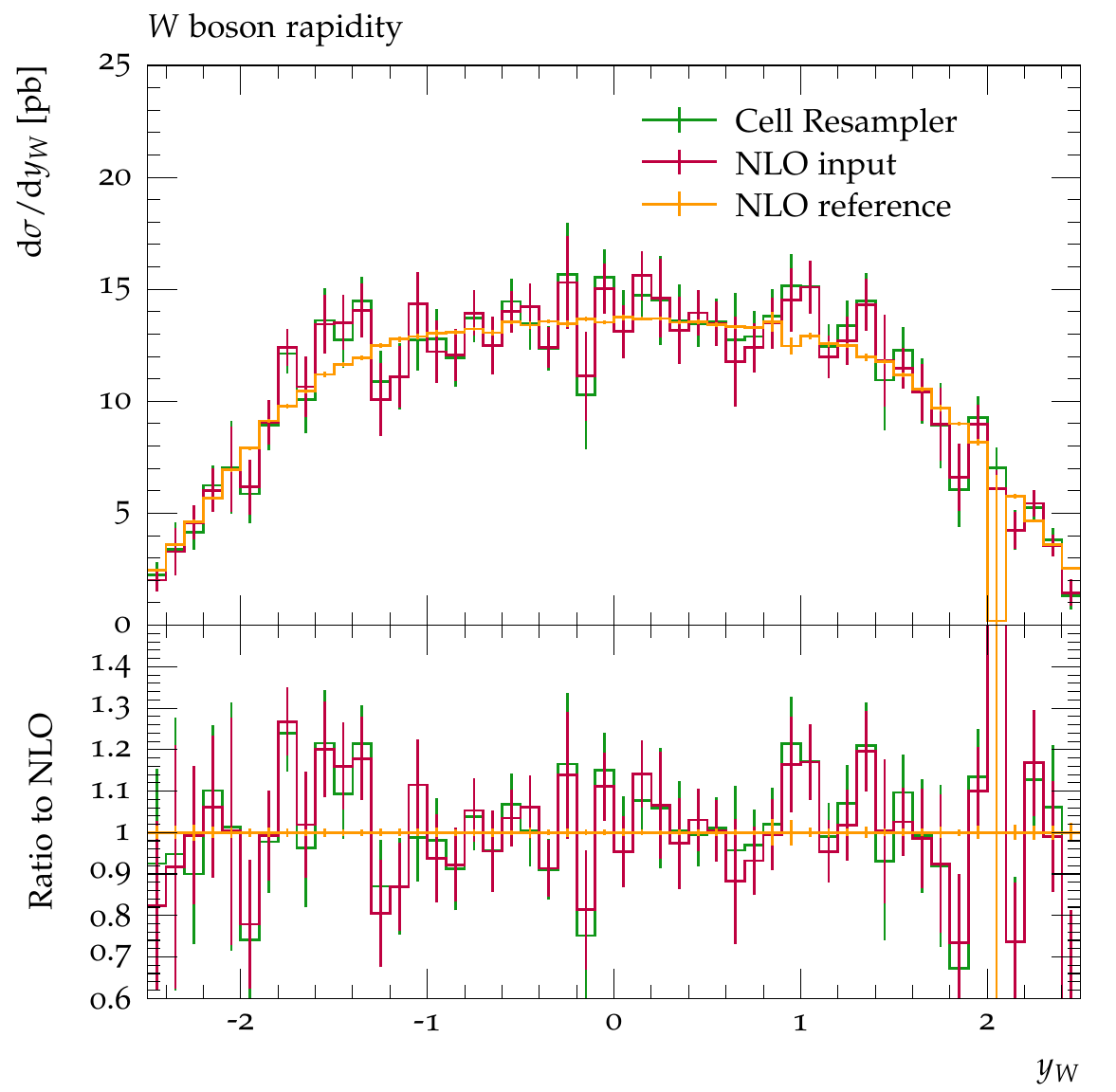}\\
    \includegraphics[width=0.44\linewidth]{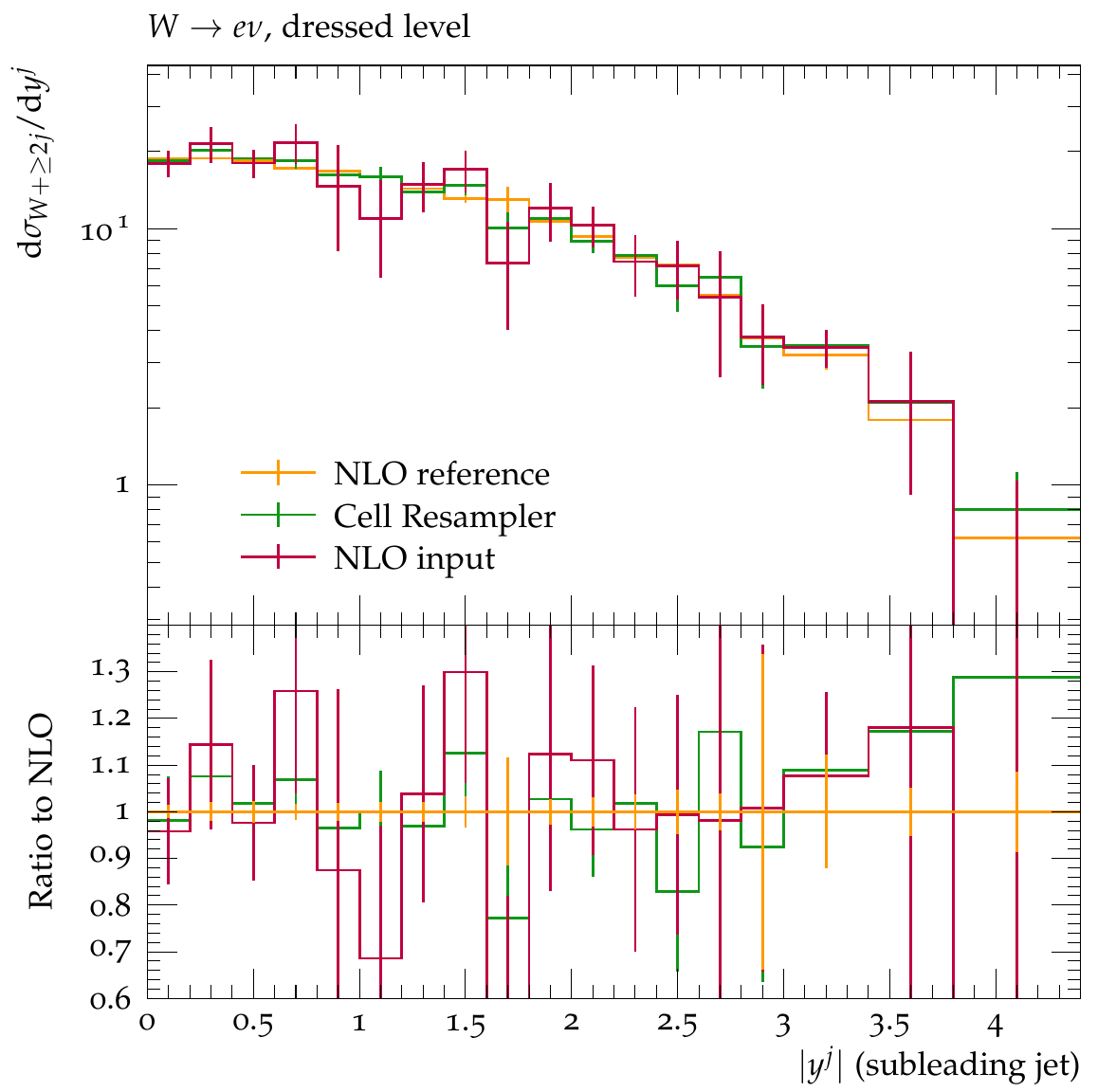}&
    \includegraphics[width=0.44\linewidth]{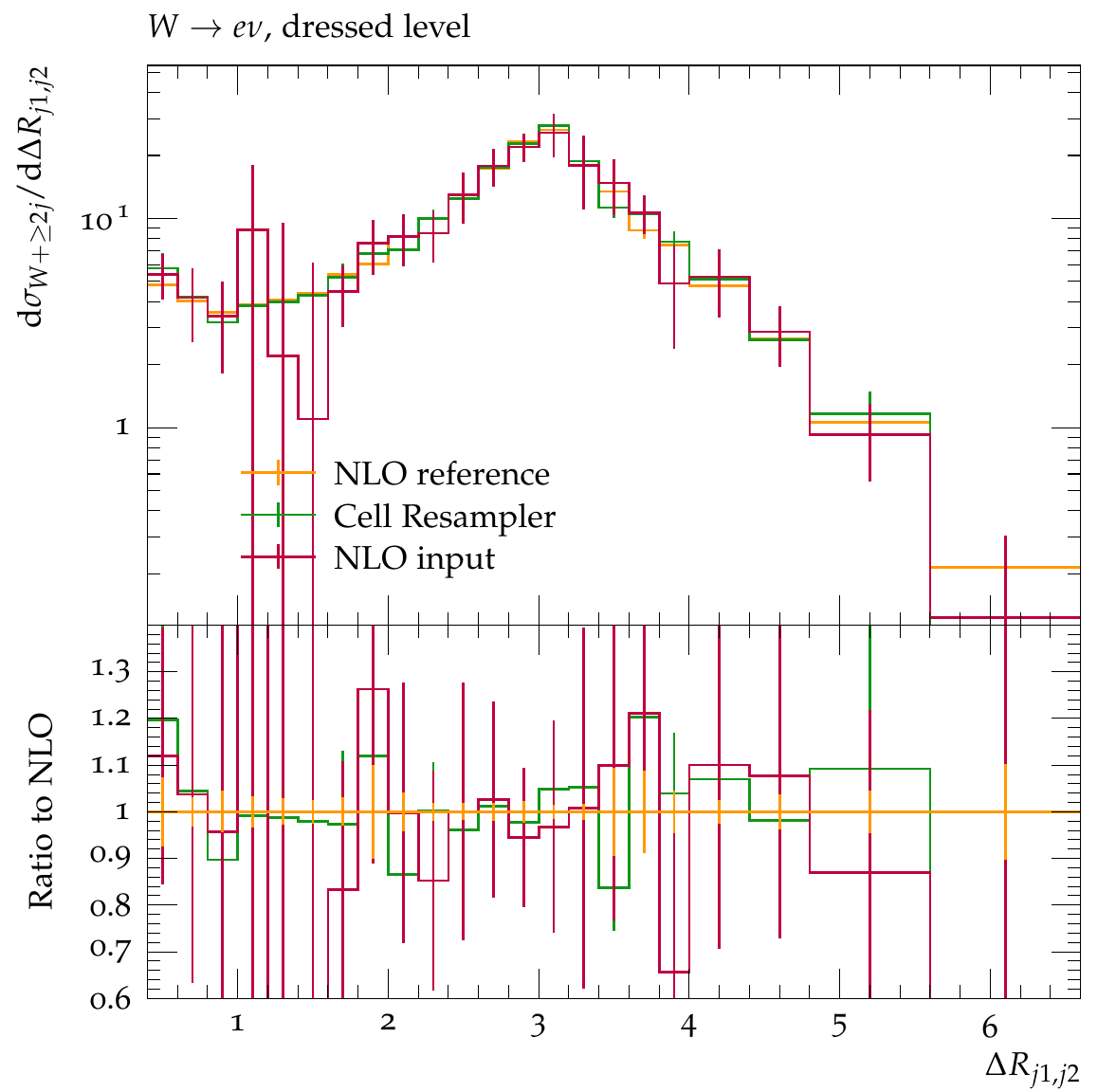}\\
    \includegraphics[width=0.44\linewidth]{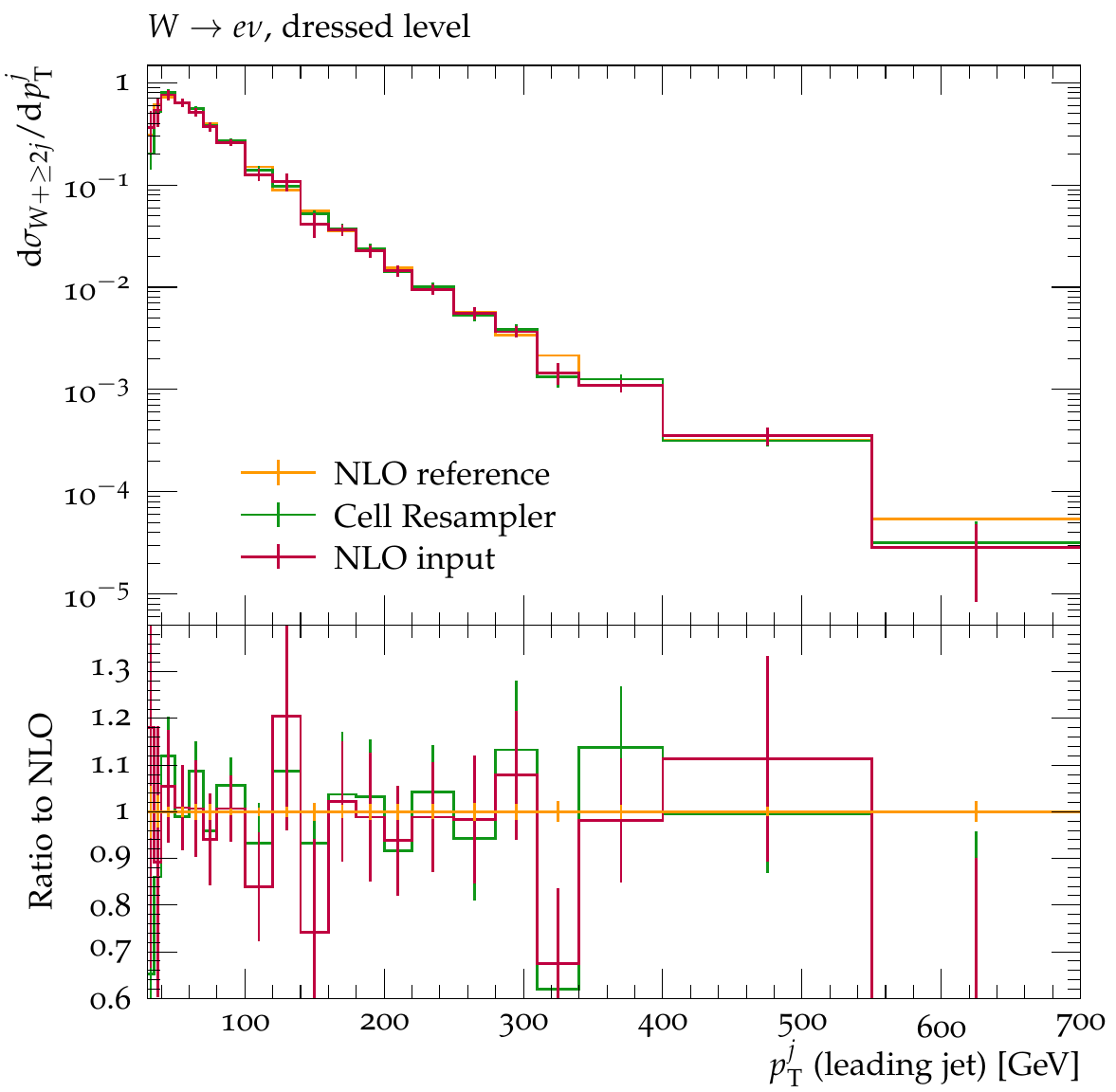}&
    \includegraphics[width=0.44\linewidth]{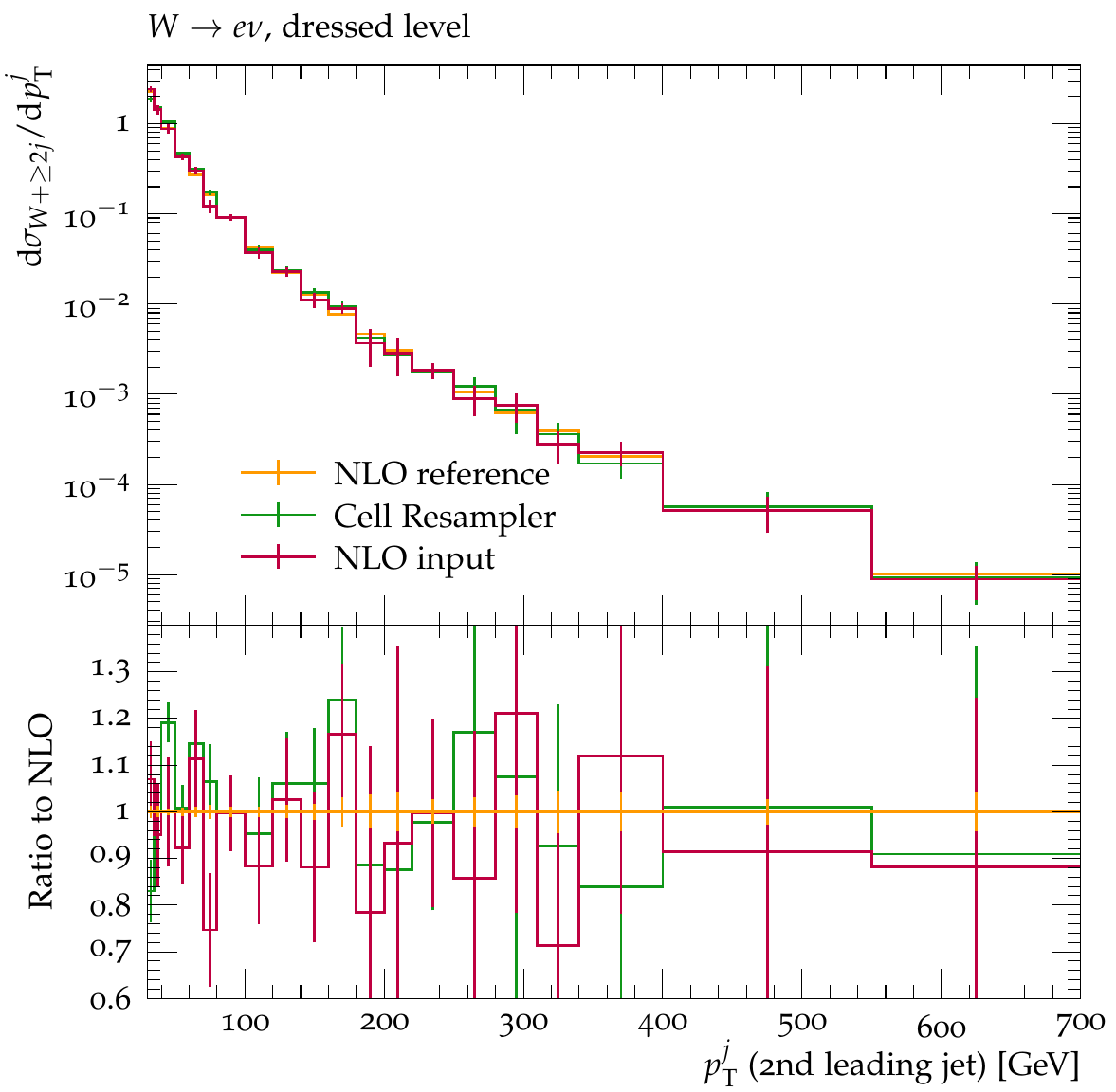}
  \end{tabular}
  \caption{The results from the pure NLO calculation and of resampler for a
    subset of the analyses performed. See text for further details.}
  \label{fig:analyses}
\end{figure}

Figure~\ref{fig:analyses} compares the vanilla prediction for the NLO
calculation and the result of passing the low-statistics NLO event
sample through the resampler for several observables. The results for
the inclusive two and three jet cross sections are shown in the top
left plot. Both the low-statistics NLO result and that of the cell
resampler are stable, as expected.
The vertical lines indicate the
statistical uncertainty estimated from the weight distribution
interpreted as a Monte Carlo sample. The estimate of the
uncertainty is larger from the pure NLO sample than that of the cell
resampler, simply because of the reduction in the variance of the
weights in the event sample.

Having demonstrated that the predictions for inclusive two and three
jet production are preserved by the cell resampling, we now consider
differential distributions. The top right plot of
figure~\ref{fig:analyses} shows the rapidity distribution of the W
boson. The level of agreement with the reference prediction within the
Monte Carlo error estimate is similar between the original
low-statistics NLO sample and the cell resampler. Concretely, the
$\chi^2$ per degree of freedom is $1.18$ for the cell resampler
and $1.25$ for the low-statistics NLO sample when
comparing to the reference distribution. We conclude that the nominal
reduction of the statistical uncertainty through cell resampling is
indeed accompanied by a stabilisation of the predicted
distributions. Furthermore, the Monte Carlo estimate of the
uncertainty in each bin correlates well with the statistical jitter
between bins.

It is important to note here that contrary to the approach
in~\cite{Andersen:2020sjs}, there is no input to the cell resampler
about any analyses intended for the sample. This implies that the
observed good agreement with the high-statistics reference predictions would not be
limited to the rapidity distribution of the W boson. Indeed, the plots
on the middle row of figure~\ref{fig:analyses} show that the reference
prediction is also matched well in the distribution in the absolute
rapidity of the hardest jet, $|y_{j_1}|$ and $\Delta R_{{j_1},{j_2}}$,
both calculated according to the analysis in~\cite{ATLAS:2014fjg}.

Finally, the bottom row in figure~\ref{fig:analyses} shows the
distributions on the transverse momentum of the hardest and second hardest
jet. These distributions are significantly smoother straight from the NLO
calculation. They are smoother still from the cell resampler. At the
same time, the cells appear to be sufficiently small, such that neither
these steeply falling distributions nor the peak in the $\Delta
R_{{j_1},{j_2}}$ distribution are smeared out to any visible degree.

\section{Conclusions}
\label{sec:conclusions}

We have presented \emph{cell resampling} as a method to eliminate
negative event weights. Negative weights are redistributed \emph{locally} in
phase space and any potential bias introduced by this redistribution
becomes arbitrarily small given sufficient statistics.

We have demonstrated the real-world performance on the highly
non-trivial example of the production of a W boson with two jets at
NLO. It is straightforward to apply our method to arbitrary processes,
and we provide an easy-to-use implementation available from
\url{https://cres.hepforge.org/} to this end.

A central ingredient of cell resampling is the definition of a
\emph{distance function} between events in phase space. This function
should mirror experimental sensitivity without referring to any
specific analysis. We have proposed a simple metric, which
is shown to perform well in practice. The exploration of more
sophisticated distance functions is a promising avenue towards future
improvements of the cell resampling method.

Since the quality of the reweighted event samples increases
systematically with their size, it is important that the computational
cost of cell resampling scales well with the number of input
events. To achieve this, we have explored an algorithm for
nearest-neighbour search in phase space based on random projections and
locality-sensitive hashing. While we find a significantly improved
scaling behaviour compared to naive linear search, further
improvements will be necessary to obtain the same increase in quality
with growing statistics.

\section*{Acknowledgements}

We thank M.~Schönherr for detailed communication about Sherpa and its
interface to \textsc{Rivet}, and D.~Maitre and T.~Kuhl for
discussions and comments on the manuscript. This work has received
funding from the European Union’s Horizon 2020 research and innovation
programme as part of the Marie Skłodowska-Curie Innovative Training
Network MCnetITN3 (grant agreement no. 722104), and the EU TMR network
SAGEX agreement No.  764850 (Marie Skłodowska-Curie).

\appendix

\section{Seed selection strategies}
\label{sec:seed_strategies}

\begin{figure}[htb]
  \centering
  \includegraphics[width=0.45\linewidth]{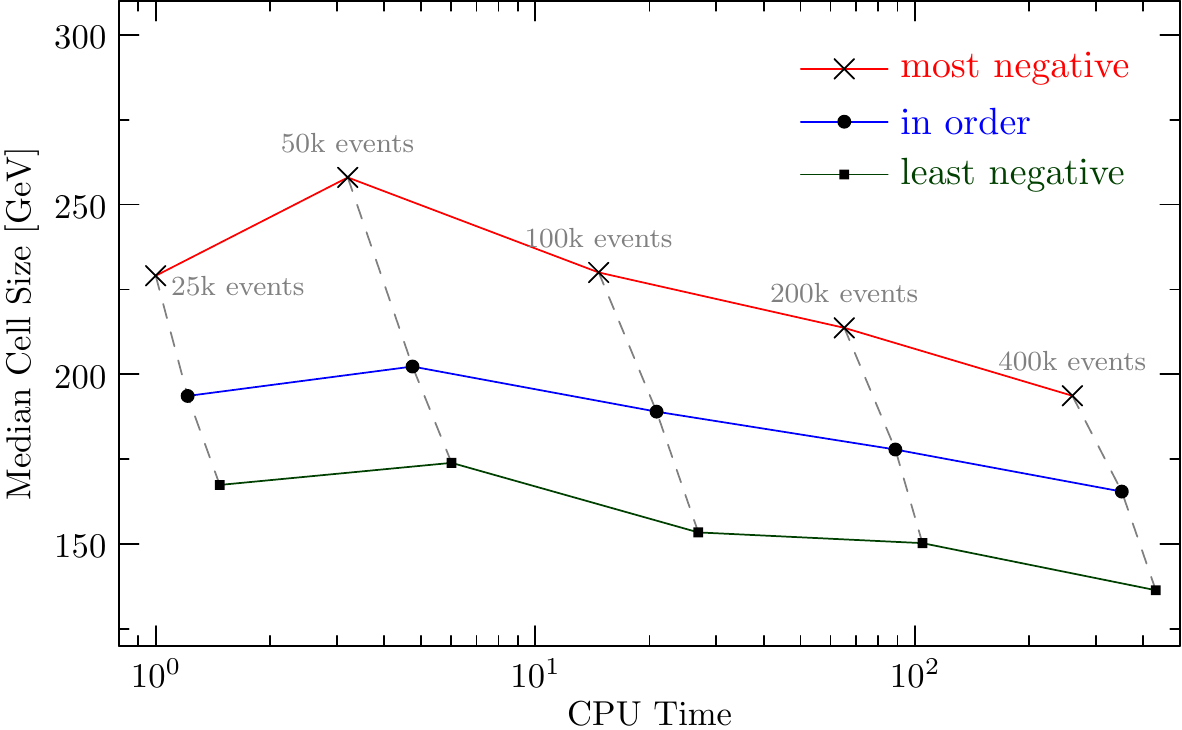}
  \includegraphics[width=0.45\linewidth]{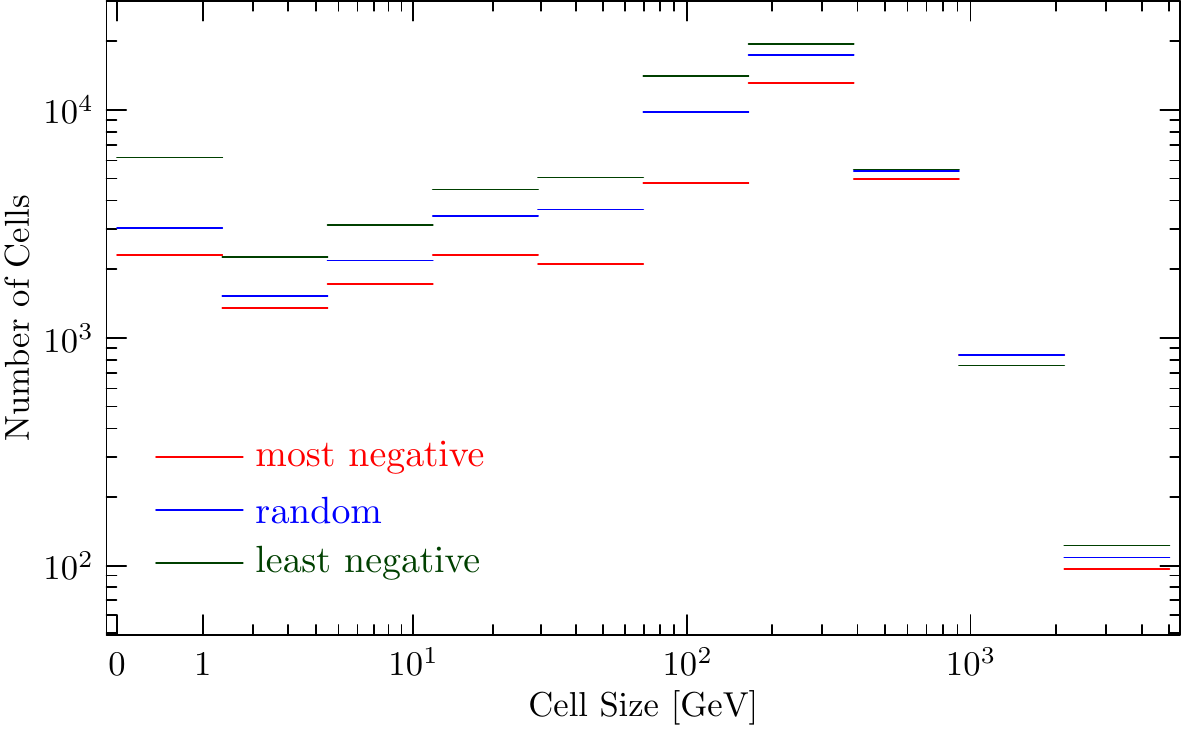}
  \caption{
    Comparison of different strategies for selecting cell seeds. On
    the left, we show the median cell radius as a function of computing
    time. Timings are normalised to the fastest run with $25\,000$ events
    and the ``most negative'' strategy. On the right, we show the
    distribution of cell radii for the sample with $400\,000$ events.
  }
  \label{fig:seed_strategies}
\end{figure}

When introducing cell resampling in section~\ref{sec:cellres}, the only
criterion for picking a cell seed was that the event weight has to be
negative. We may wonder whether constructing cells in a specific order
changes the final result. To answer this question, we assess three
different strategies for selecting the seed for the next cell.
\begin{enumerate}
\item Choose the event with the most negative weight (``most negative'').
\item Choose the next negative-weight event according to the order in which events were generated (``in order'').
\item Choose the negative-weight event with weight closest to zero (``least negative'').
\end{enumerate}
We are primarily interested in the quality of the outcome, i.e.~the
resampling should affect observables as little as possible. As a proxy
for this very general requirement, we first consider the cell size
distributions. After quality, a secondary criterion is speed. In
figure~\ref{fig:seed_strategies}, we compare the median cell sizes as
a function of computing time for samples with between 25 and 400
thousand events. We also show the distribution of cell radii for the
largest event sample.
We might be tempted to conclude that the ``least negative'' strategy
is best: despite being the slowest for any given sample, it
consistently produces smaller cells given either a fixed number of
input events or even a fixed computing time budget. However, we also
notice that this strategy not only produces more small-sized cells,
but also a larger number of big cells. By choosing seeds with weight
close to zero first, we tend to construct cells with increasing radii.
Cells that are constructed early on may easily be subsumed by later
cells, which implies that the time spend on constructing the earlier
cell is wasted and the median cell size is no longer a good indicator
of quality.

We can limit the impact of big cells by imposing a maximum cell
radius. In this case, we can measure the quality of the resampling
through the negative-weight factor
\begin{equation}
  \label{eq:sigma_-}
  \hat{\sigma}_- = \frac{1}{\sigma} \frac{d\sigma}{d\sgn w}\bigg|_{w=-1} = -\frac{\sum_{w_i <0} w_i}{\sum_{i} w_i},
\end{equation}
where $w_i$ is the weight of event $i$. For the initial event sample
with $400\,000$ events we find $\hat{\sigma}_- \approx
26.1$. Resampling with a maximum cell radius of $100\,$GeV yields
$\hat{\sigma}_- \approx 1.1$, with fluctuations at the level of one
per cent between the different strategies. We do not observe any
significant differences in run time.

In conclusion, we propose the following modus operandi. First, we
estimate the typical cell size through test runs without imposing any
limit on the radius. To save computing time we can employ the ``most
negative'' strategy and extrapolate from smaller event samples. We
then perform the actual resampling with a cell radius limit of the
order of the estimated median. Following the above discussion, we do
not expect any significant dependence of the final results on the cell
selection strategy.

\bibliographystyle{JHEP}
\bibliography{papers}

\end{document}